\documentclass[letterpaper,11pt]{report}

\pdfoutput=1

\title
{
	Managing Distributed MARF with SNMP
}

{\author
	{
		Serguei A. Mokhov\\
		Lee Wei ``Lewis'' Huynh\\
		Jian ``James'' Li\\\\\\\\\\\\
		Concordia Univerisity\\
		Montr\'eal, Qu\'ebec, Canada\\\\\\
	}
}

\date{Tue  2 Jun 2009 05:22:48 EDT
}

\usepackage{graphicx}
\usepackage{latexsym}
\usepackage{makeidx}
\usepackage{url}
\usepackage{hyperref}

\makeindex

\newcommand{\marf}{\textsf{MARF}}

\newcommand{\xf}[1]{Figure~\ref{#1}}

\newcommand{\xs}[1]{Section~\ref{#1}}

\newcommand{\file}[1]{\texttt{#1}\index{Files!#1}}

\newcommand{\api}[1]{\texttt{#1}\index{API!#1}}

\begin{document}

	\begin{titlepage}
		\maketitle
	\end{titlepage}

	\pagenumbering{roman}
\tableofcontents
\clearpage
\pagenumbering{arabic}

\listoffigures

	\chapter{Introduction}
\index{Introduction}

$Revision: 1.1.2.6 $

\section{Background}

The Modular Audio Recognition Framework ({\marf})~\cite{marf,marf02,dmarf06} is an open-source research
platform and a collection of voice, sound, speech, text,  and natural language
processing (NLP) algorithms written in Java and arranged into a modular
and extensible framework facilitating addition of new algorithms. MARF can
run distributedly over the network (using CORBA, XML-RPC, or Java RMI) and
may act as a library in applications or be used as a source for learning
and extension. A few example applications are provided to show how to use
the framework. One of MARF's applications, \api{SpeakerIdentApp} has a database
of speakers, where it can identify who people are regardless what they
say.

Original {\marf}~\cite{marf02} was developed by Serguei Mokhov with a few classmates and
others throughout a variety of courses. Distributed MARF~\cite{dmarf06} (DMARF) proof-of-concept (PoC)
implementation was done by Serguei Mokhov in the Distributed Systems class.
The Distributed MARF nodes are hard to manage if there are many, including
configuration, statistics, and status management.

{\marf} has several applications. Most revolve around its recognition
pipeline -- sample loading, preprocessing, feature extraction, and training or classification.
One of the applications, for example is Text-Independent Speaker
Identification. In the classical MARF, the pipeline and the applications as they stand are purely
sequential with even little or no concurrency when processing a bulk of voice
samples. Thus, the purpose of DMARF in~\cite{dmarf06} was to make the pipeline distributed and run on a cluster
or a just a set of distinct computers to compare with the traditional version and add disaster recovery and service replication,
communication technology independence, and so on.

\begin{figure}
	\centering
	\includegraphics[width=\textwidth]{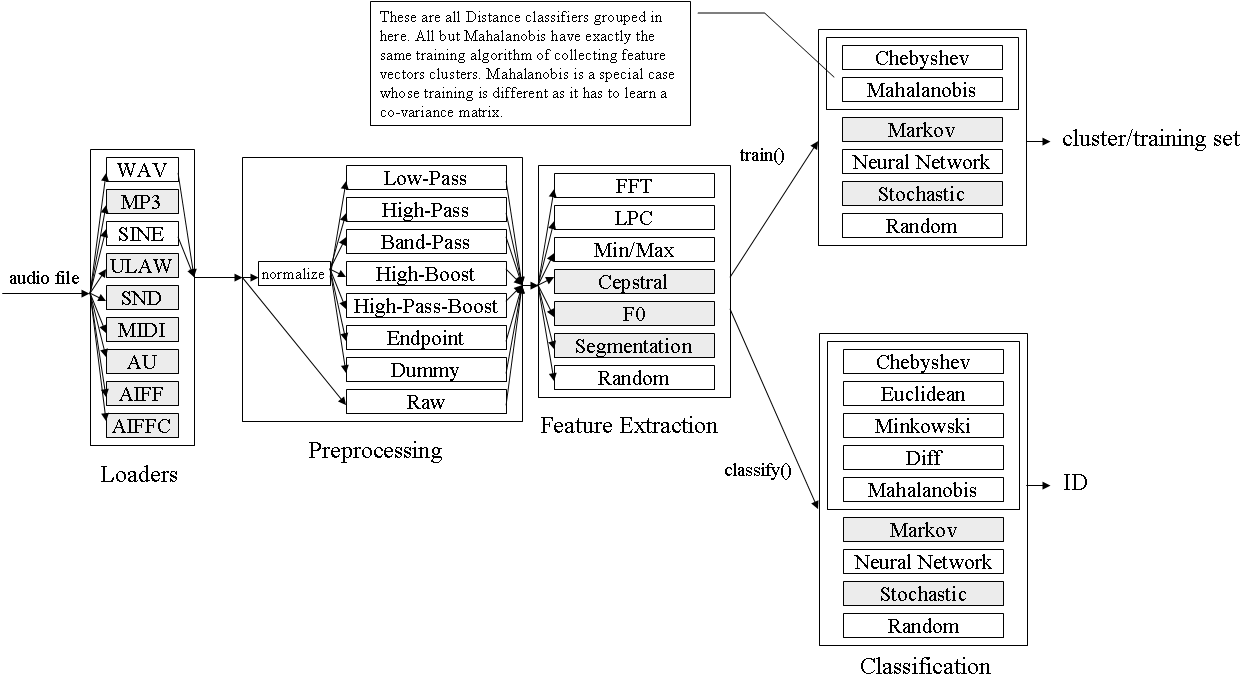}
	\caption{The Core MARF Pipeline\index{MARF!Core Pipeline} Data Flow}
	\label{fig:pipeline-flow}
\end{figure}

The classical {\marf}'s pipeline is in \xf{fig:pipeline-flow}.
The goal of DMARF was to distribute the shown stages of the pipeline as
services as well as stages that are not directly present in
the figure -- sample loading, front-end application service (e.g.
speaker identification service, etc.) among other things
in the distributed system. The reasons to be able flexibly distribute
these services is to offload the bulk of multimedia/data crunching and processing to a
higher performance servers, that can communicate, while the data collection
may happened at several low-cost computers (e.g. low-end laptops) or PDAs alike,
embedded devices, etc., which may not necessarily have the processing power and
storage capacity locally to cope with the amount of incoming data, so they pass it on
to the servers. (A possible infrastructure of such a setup can, for example,
be in place in different law enforcement agencies spread out across a country,
yet, being able to identify speakers across all jurisdictions if say recorded phone
conversations of a suspect are available. In another scenario, the could be
used in tele-conferencing.)

\begin{figure}
	\centering
	\includegraphics[width=\textwidth]{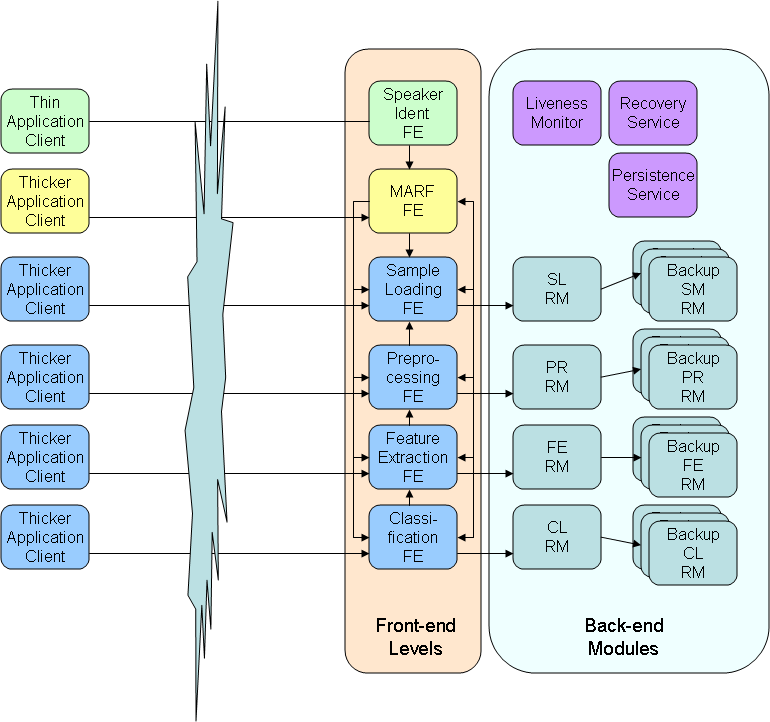}
	\caption{The Distributed MARF Pipeline\index{MARF!Distributed Pipeline}}
	\label{fig:pipeline-net}
\end{figure}

\clearpage

In \xf{fig:pipeline-net} the distributed version of the pipeline is
presented. It indicates different levels of basic front-ends, from
higher to lower, which a client application may invoke as well
as services may invoke other services through their front-ends
while executing in a pipeline-mode. The back-ends are in charge
of providing the actual servant implementations as well as the
features like primary-backup replication, monitoring, and disaster
recovery modules.

The status
management graphical user interface (GUI) prototypes were developed in DMARF, but not implemented
to provide application managers an ability to monitor services' status (or the whole pipeline)
as well as change configuration options, as in \xf{fig:speak-gui-client} and in \xf{fig:server-status-gui}.

\begin{figure}[hb]
	\includegraphics[totalheight=\textwidth,angle=90]{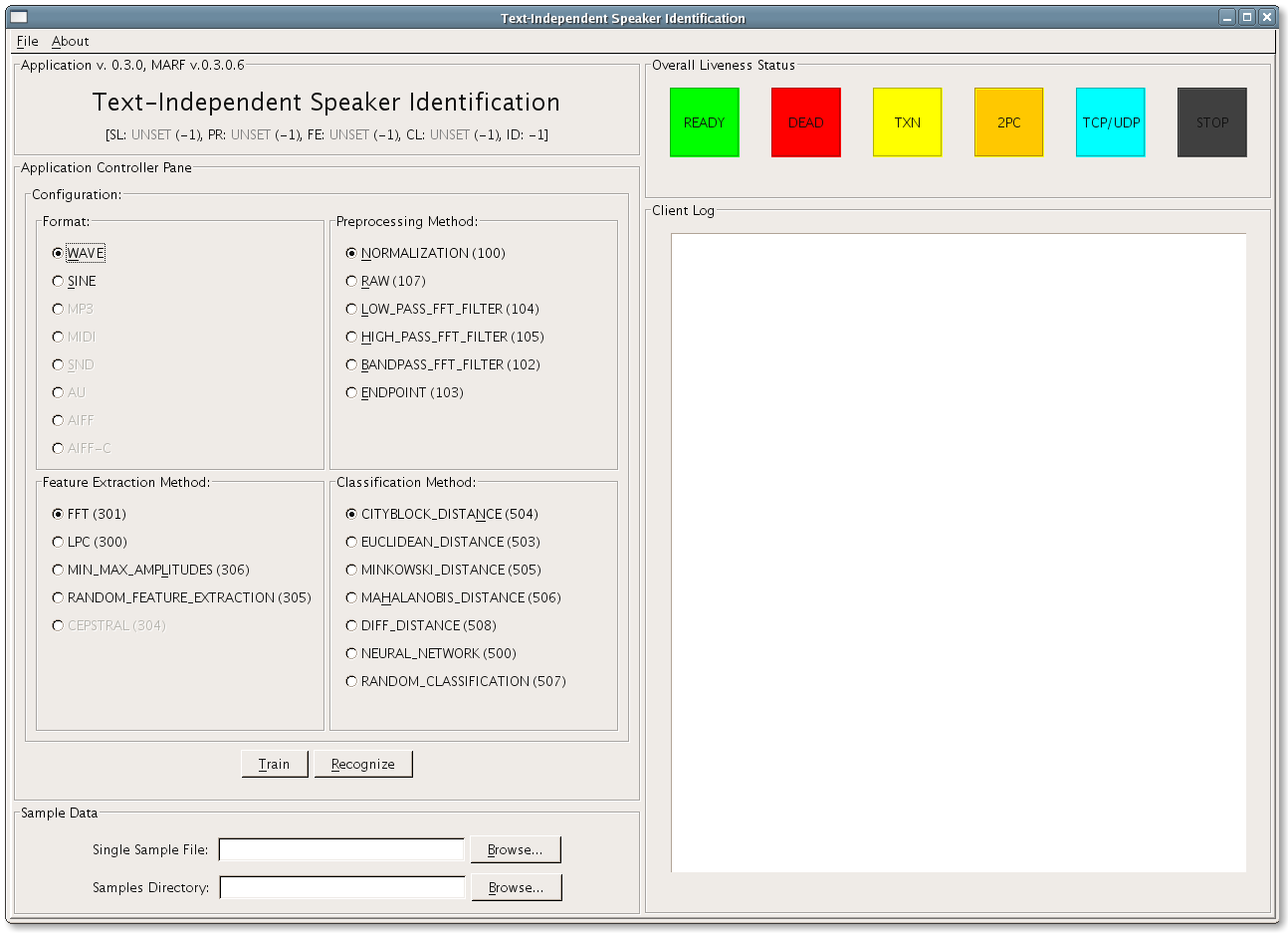}
	\caption{SpeakerIdenApp Client GUI Prototype (Manager)}
	\label{fig:speak-gui-client}
\end{figure}

\begin{figure}[hb]
	\includegraphics[width=\textwidth]{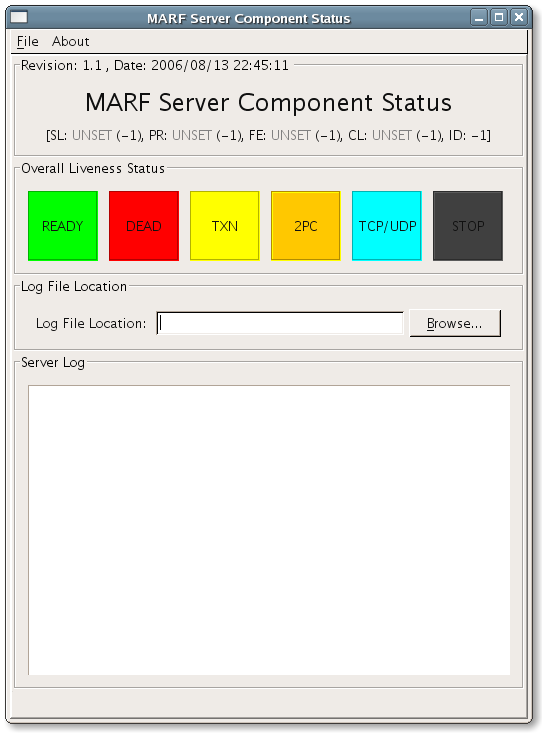}
	\caption{MARF Service Status Monitor GUI Prototype (Agent)}
	\label{fig:server-status-gui}
\end{figure}

\section{Scope}

The scope of this project's work focuses on the research and prototyping of
the extension of the Distributed MARF such that its services
can be managed through the most popular management protocol
familiarly, SNMP. The rationale behind SNMP vs. MARF's proprietary management
protocols, is that can be
integrated with the use of common network service and device
management, so the administrators can manage MARF nodes via a
already familiar protocol, as well as monitor their performance,
gather statistics, set desired configuration, etc. perhaps using
the same management tools they've been using for other
network devices and application servers.

MARF has generally thee following type of services: application, core pipeline, sample loading, preprocessing,
feature extraction, and classification. There are common data structures, configuration, storage management
attributed to them. DMARF's components in general are stand-alone and may listen on RMI, XML-RPC, CORBA,
and TCP connections for their needs, and natively do not ``understand'' SNMP. Therefore, each managed service
will have to have a proxy SNMP-aware agent for management tasks and delegate instrumentation proxy to
communicate with the service's specifics.
Thus, in this work we are designing and implementing to some extent the following:

\begin{itemize}
\item
Defining MIBs for MARF Services
\item
Producing Proxy SNMP Agents
\item
Agent-Adapter Delegate Instrumentation
\item
SNMP MARF Manager Applications
\end{itemize}

The original proposal has a lot more provisional tasks, that are outlined in
\xs{sect:future-work}.

\section{Tools}

We will use platform-independent tools like Java, possibly
JDMK based on JMX, eventually RMI, CORBA, XML-RPC as provided by AdventNet or others.
SimpleWeb \cite{simpleweb-mibvalidation} for original MIB cross-validation
and AdventNet's SNMP Java API \cite{advent-net-java-api} and Java Agent SDK \cite{advent-net-java-agent-sdk} tools
are in actual active use.

\section{Summary}

The project seems like a viable and useful option to extend MARF and
contribute its implementation at the same time to the open-source
community.


	\chapter{Methodology}
\index{Methodology}

$Revision: 1.1.2.10 $

\section{Introduction}

Distributed MARF \cite{marf} offers a number of service types:

\begin{enumerate}
\item
Application Services
\item
General MARF Pipeline Services
\item
Sample Loading Services
\item
Preprocessing Services
\item
Feature Extraction Services
\item
Training and Classification Services
\end{enumerate}

\noindent
which are backed by the corresponding server implementations
in CORBA, Java RMI, and Web Services XML-RPC. The services
can potentially be embedded into other application or hardware
systems for speaker and language identification, and others.

We are interested in managing such services over a network as a whole,
collect their processing statistics, perform remote management, etc., so we need to
define them in MIB using ASN.1 including the types of requests
and responses and their stats.

We have applied to IANA \cite{iana-pen} and registered a private enterprise number (PEN)
under the \api{enterprises} node, which is 28218 (under the MARF Research and Development Group).

\section{MARF-Manager-Agent Architecture}

\begin{figure}
	\centering
	\includegraphics[width=.9\textheight,angle=90]{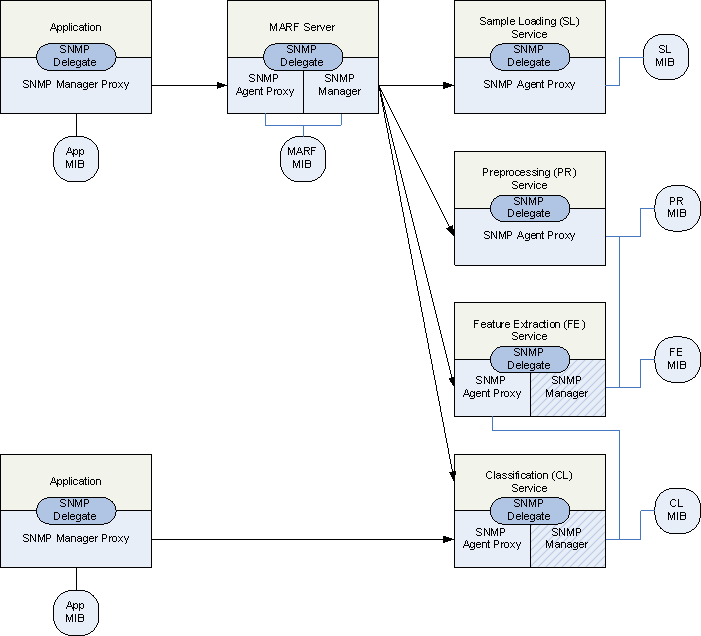}
	\caption{MARF-Manager-Agent Architecture}
	\label{fig:marf-managers-agents-proxies-arch}
\end{figure}

We devised a preliminary management architecture for MARF application
and services, presented in \xf{fig:marf-managers-agents-proxies-arch}.
In this figure we capture a relationship between all the major entities
in the system. Application are the ultimate managers, whereas, the remaining
services can be both managers and agents in some cases. The MARF service
which operates the pipeline on behalf of main applications, can manage
sample loading, preprocessing, feature extraction, and classification.
Since those services can talk to each other and request data from each
other (e.g. classification can request data from feature extraction),
they may exhibit manager characteristics, not fully explored in this work.
Applications don't need to come through the MARF service manager, but in
case some need arises (debugging, development, maintenance), can connect
to the terminal services of the pipeline directly.

\section{SMI Structure}

\begin{figure}
	\centering
	\includegraphics[width=0.9\textheight,angle=90]{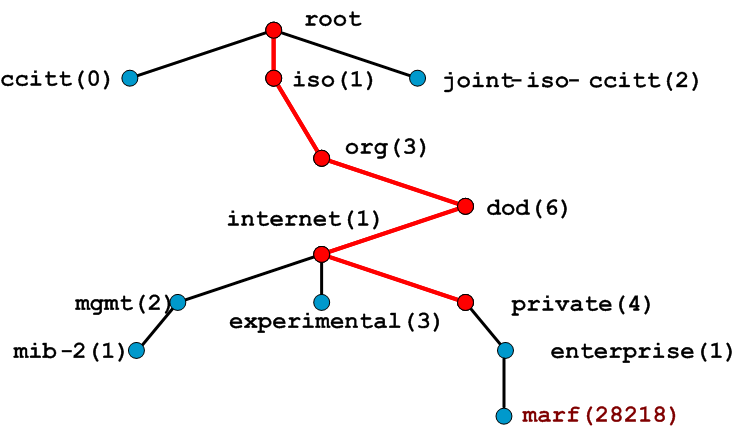}
	\caption{Preliminary MARF Private Enterprises Number.}
	\label{fig:enterprises-marf-mib}
\end{figure}

\begin{figure}
	\centering
	\includegraphics[width=0.9\textheight,angle=90]{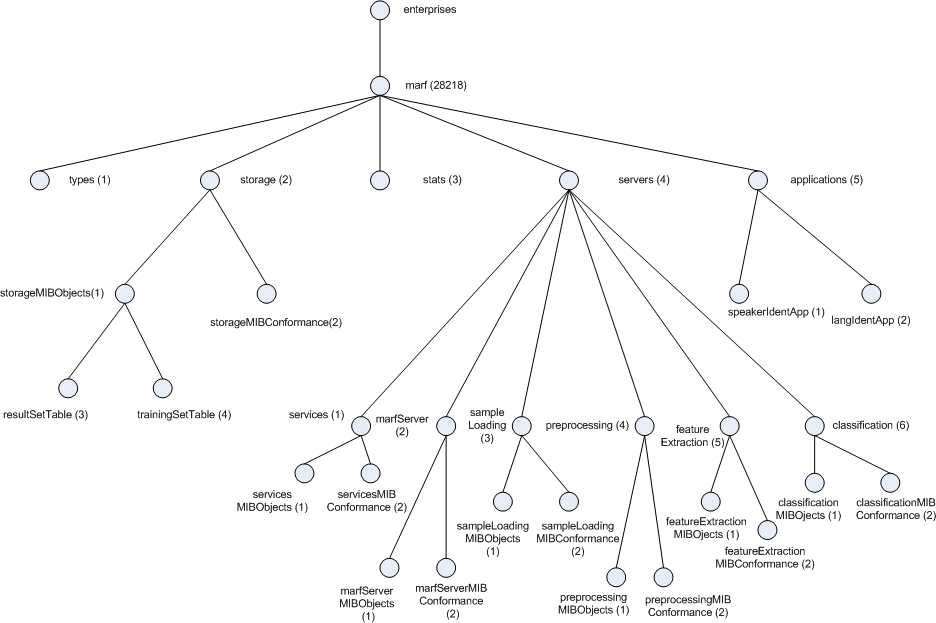}
	\caption{Preliminary MARF General Tree.}
	\label{fig:marf-mib}
\end{figure}

We have worked on the MIB tree for the entire DMARF and did some progress
in defining general services and storage types, Preprocessing,
Feature Extraction, Classification, SpeakerIdentApp, and LangIdentApp MIBs,
which are under the \file{marf/src/mib} directory.
We started off with \file{WWW-MIB} \cite{www-mib} as an example for a \api{WwwService}'s
and some works from \file{ATM-TC-MIB} and related files from the same source.
We preliminary completed the indicated MIBs and provided default proxy implementations for the
services, at least a few of them using the AdventNet's API and SDK, which can be
found under \file{marf/src/marf/net/snmp} directory tree. In \xf{fig:enterprises-marf-mib} we show
where MARF's MIB subtree begins under the \api{enterprises} node.
And in \xf{fig:marf-mib} we present the general overview of the most SMI
components for MARF services and other required components. We also
used lecture notes of Dr. Chadi Assi \cite{assi-inse7120-2007}.
We primarily using SNMPv2 along with SMIv2 in this project.

\section{MARF Services}

This section provides a sneakpeek on the MARF service types we are dealing
with and some of them have their preliminary MIB subtrees drawn. As mentioned
earlier we provide our MIBs so far in \file{marf/src/mib}. Note, some of the
diagrams may not correspond 1-to-1 to MIBs as it is work in progress and things
sometimes get out of sync.
The relevant files are (as of this writing):

\begin{itemize}
\item
\file{MARF-MIB.mib} -- main file meant to consolidate all when ready
\item
\file{MARF-types.mib} -- some common textual conventions
\item
\file{MARF-storage.mib} -- storage-related issues and types so far
\item
\file{MARF-services.mib} -- general services description
\item
\file{MARF-sample-loading.mib} -- concrete sample loading service
\item
\file{MARF-preprocessing.mib} -- concrete preprocessing service
\item
\file{MARF-feature-extraction.mib} -- concrete feature extraction service
\item
\file{MARF-classification.mib} -- concrete classification service
\item
\file{MARF-APPS-SPEAKERIDENTAPP.mib} -- a MIB for SpeakerIdentApp
\item
\file{MARF-APPS-LANGIDENTAPP.mib} -- a MIB for LangIdentApp
\end{itemize}

\subsection{General Service MIB}

Most MARF services share some common definitions about indexing the services,
their statistics, and so on, so the general service module was created such
that most of this general functionality is captured in there including statistics, and the more
specific modules extend by augmenting its tables and using its types. The generic
service description is in \xf{fig:services-mib}.

\begin{figure}
	\centering
	\includegraphics[width=.9\textheight,angle=90]{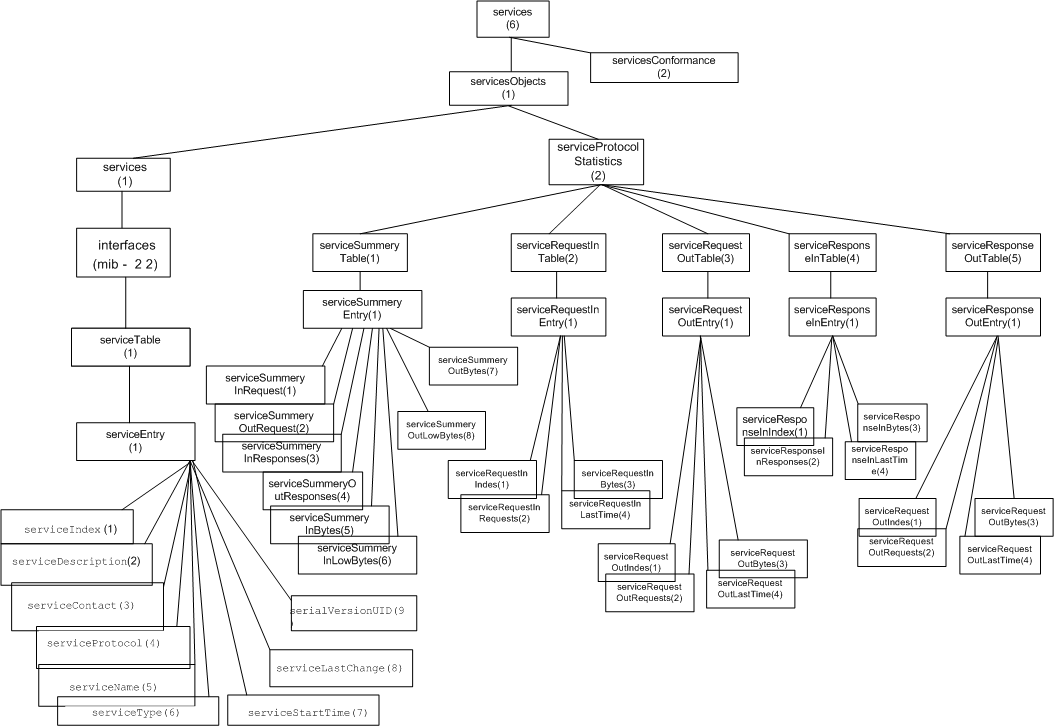}
	\caption{General Service MIB 1.}
	\label{fig:services-mib}
\end{figure}

\begin{figure}
	\centering
	\includegraphics[width=.9\textheight,angle=90]{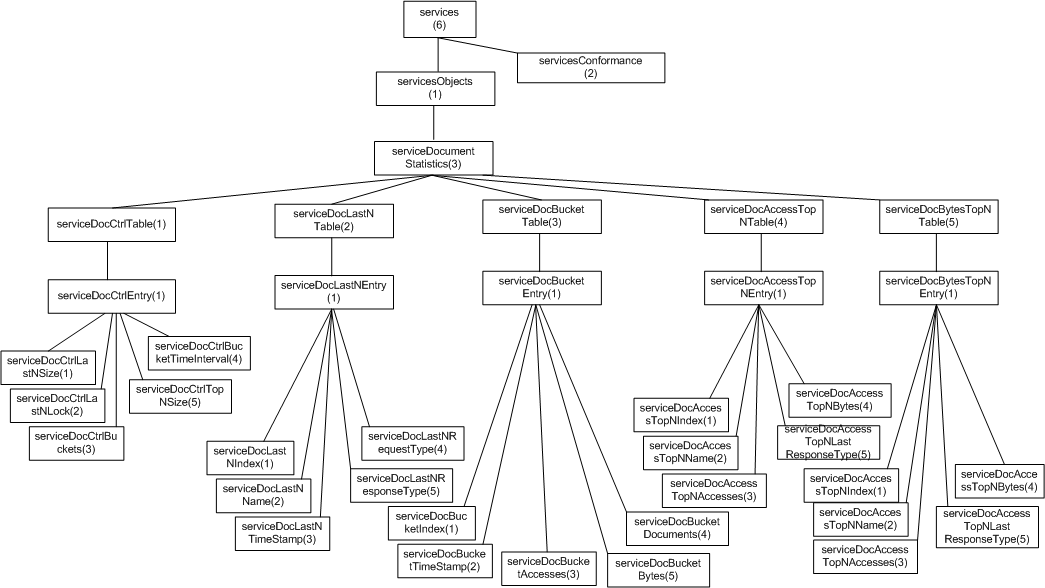}
	\caption{General Service MIB 2.}
	\label{fig:services-mib}
\end{figure}

\subsection{Storage}

MIB for storage related activities (e.g. training sets, classification results, etc.)
has to be provided, as such the MIB presented in \xf{fig:storage-mib} was devised.

\begin{figure}
	\centering
	\includegraphics[width=\textwidth]{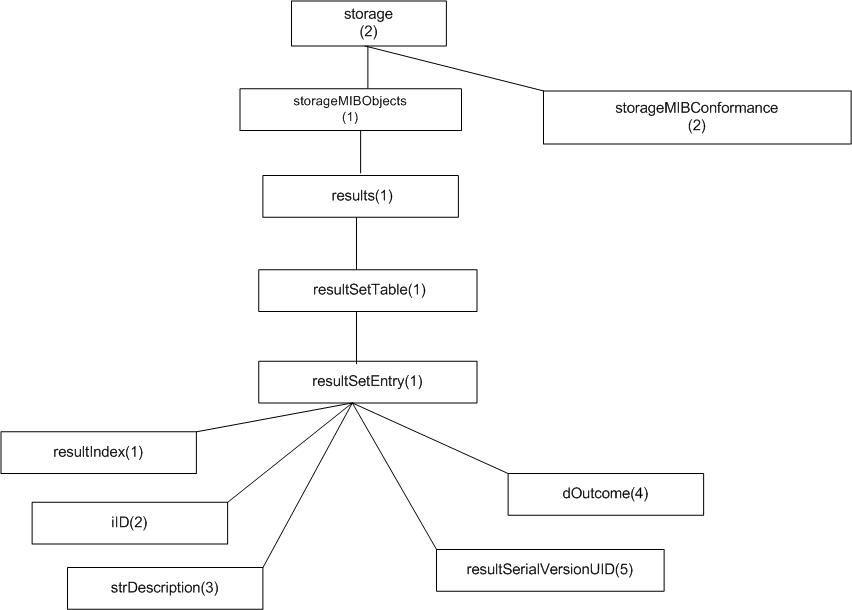}
	\caption{Storage MIB.}
	\label{fig:storage-mib}
\end{figure}

\subsection{Sample Loading}

The Sample Loading Service knows how to load certain file or stream types (e.g. WAVE) and convert them accordingly for further preprocessing. In our project, we are introducing the sample loading module, which we will use to manage and keep tracking its specific parameters which are:

\begin{enumerate}
\item
\api{iFormat}: a sample format, an integer type attribute.
\item
\api{adSample}: sample data, a collection of bytes.
\end{enumerate}

All these attributes will be located in \api{sampleLoadingServiceEntry} object which is in \api{sampleLoadingServiceTable} object.
See the example of the SMI tree for sample loading in \xf{fig:sample-loading-mib}.

\begin{figure}
	\centering
	\includegraphics[width=.9\textwidth]{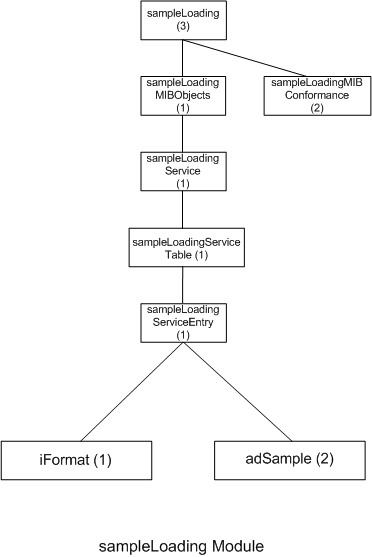}
	\caption{Preliminary MARF Sample Loading Service MIB.}
	\label{fig:sample-loading-mib}
\end{figure}

\subsection{Preprocessing}

The Preprocessing service accepts incoming voice or text samples and does the requested preprocessing (all sorts of filters, normalization, etc.). Its function generally is to normalize the incoming sample file by amplitude starting from certain index and/or filter it. It has several algorithms to be as an option to filter the voice frequencies. The algorithms in this module are FFT-based and CFE-based. In our project, we introduce the preprocessing MIB module, which will use to manage and keep tracking its specific parameters which are:

\begin{enumerate}
\item
\api{Sample}: a collection of doubles and a format.
\item
\api{dSilenceThreshold} : a double type for the silence cut off threshold.
\item
\api{bRemoveNoise} : a Boolean type to indicate noise removal.
\item
\api{bRemoveSilence} : a Boolean type to indicate silence removal.
\end{enumerate}

All these attributes are located in the \api{preprocessingServiceEntry} object which is in
\api{preprocessingServiceTable} tabular object.
See the example of the SMI tree for preprocessing in \xf{fig:preprocessing-mib}.

\begin{figure}
	\centering
	\includegraphics[width=\textwidth]{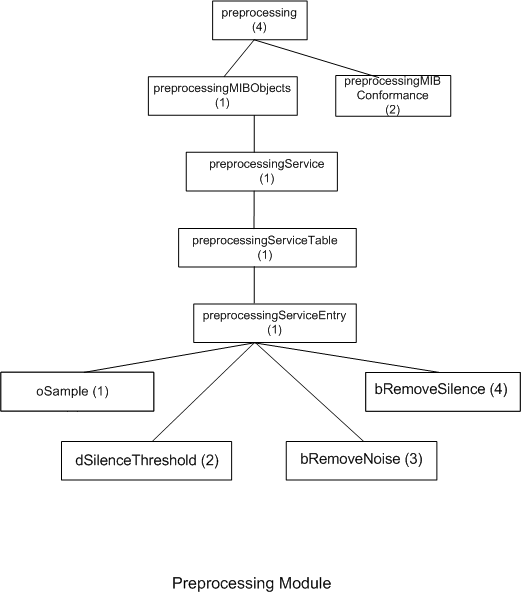}
	\caption{Preliminary MARF Preprocessing Service MIB.}
	\label{fig:preprocessing-mib}
\end{figure}

\subsection{Feature Extraction}

Feature Extraction service accepts data, presumably preprocessed, and attempts to extract features out of it given requested algorithm (out of currently implemented, like FFT, LPC, MinMax, etc.) and may optionally query the preprocessed data from the Preprocessing Service.
See the SMI tree for feature extraction in \xf{fig:feature-extraction-mib}.
The parameters been used in the Feature Extraction are below:

\begin{enumerate}
\item
\api{adFeatures}: a currently being processed feature vector.
\item
\api{oFeatureSet}: a collection of feature vectors or sets for a subject.
\end{enumerate}

\begin{figure}
	\centering
	\includegraphics[totalheight=\textwidth,angle=90]{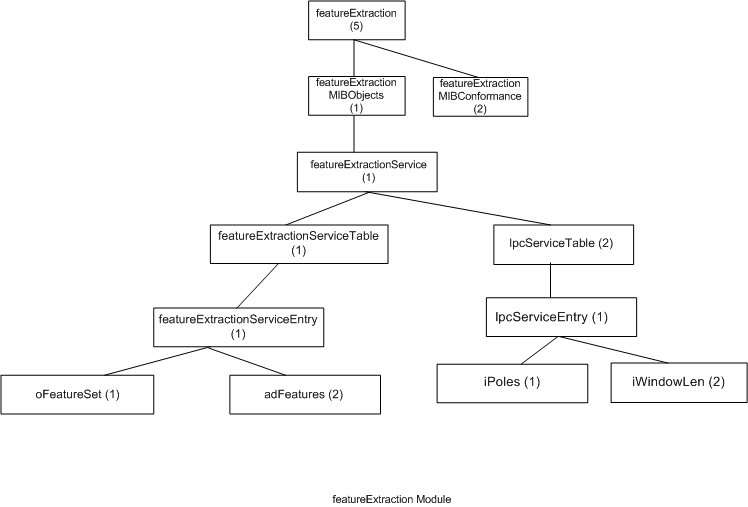}
	\caption{Preliminary MARF Feature Extraction Service MIB.}
	\label{fig:feature-extraction-mib}
\end{figure}

\subsection{Classification}
\index{Classification}

Classification-type of services are responsible for either training
on features data (read/write) or classification (read-only) of the
incoming data. Being a bit similar in operation to the service types
mentioned earlier, Classification does a lot more legwork and is actually
responsible saving the training set database (in files or otherwise) locally,
provide concurrency control, persistence, ACID, etc. Its MIB tree
in \xf{fig:classification-mib} is very similar to the other services
at this point.

\begin{figure}
	\centering
	\includegraphics[width=\textwidth]{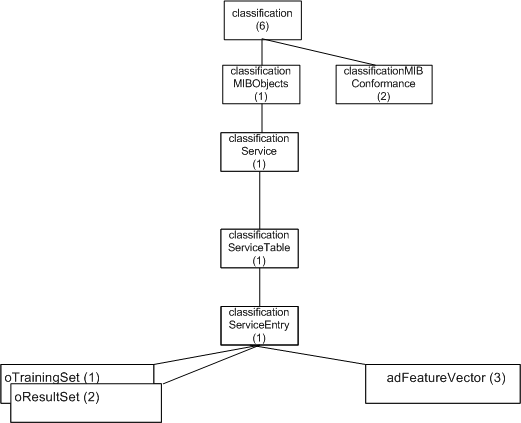}
	\caption{Preliminary MARF Classification Service MIB.}
	\label{fig:classification-mib}
\end{figure}

\begin{enumerate}
\item
\api{adFeatures}: a currently being processed feature vector for training/classification.
\item
\api{oResultSet}: a collection of classification results.
\end{enumerate}

\clearpage

\subsection{Applications}
\index{Applications}

MARF has many applications, for this project we were considering
primarily two applications for management, SpeakerIdentApp as well
as LangIdentApp for speaker and language identification.
The MIB trees we designed are in
in \xf{fig:SPEAKERIDENTAPP-MIB} and \xf{fig:langidentapp-mib}.

\begin{figure}
	\centering
	\includegraphics[width=\textwidth]{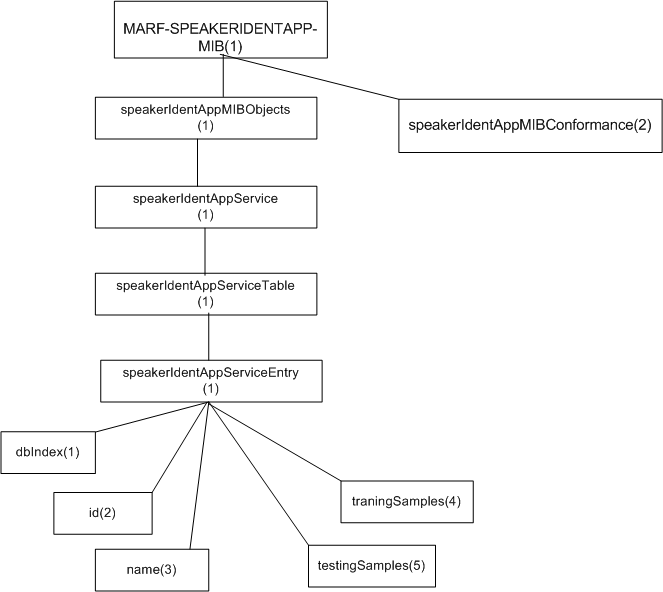}
	\caption{SpeakerIdentApp MIB.}
	\label{fig:SPEAKERIDENTAPP-MIB}
\end{figure}

\begin{figure}
	\centering
	\includegraphics[width=\textwidth]{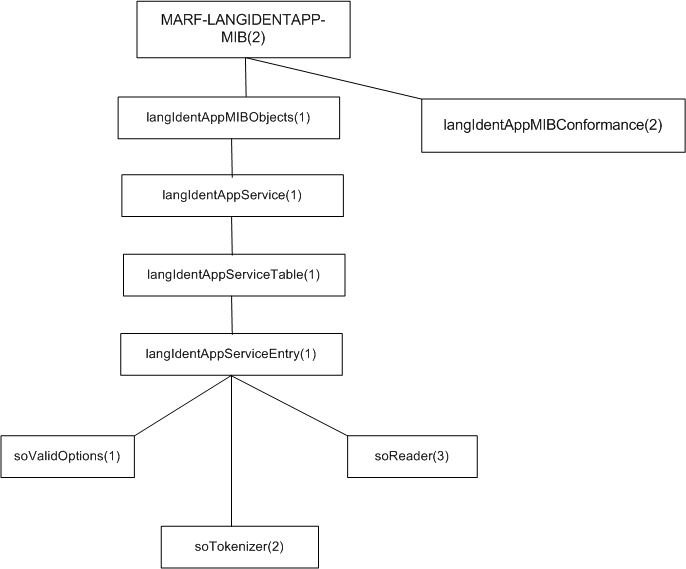}
	\caption{LangIdentApp MIB.}
	\label{fig:langidentapp-mib}
\end{figure}


	\chapter{Conclusion}

$Revision: 1.1.2.5 $

\section{Review of Results}

\subsection{MIBs}

By far, we have successfully finished a set of compilable and loadable MIBs
from different aspects of DMARF over SNMP:

\begin{enumerate}
\item
MARF-MIB
\item
MARF-types
\item
MARF-storage
\item
MARF-services
\item
MARF-sample-loading
\item
MARF-preprocessing
\item
MARF-feature-extraction
\item
MARF-classification
\item
MARF-APPS-SPEAKERIDENTAPP
\item
MARF-APPS-LANGIDENTAPP
\end{enumerate}

While some of the information in the above MIBs is in debug
form, it is enough to compile and generate proxy agent for
minimal testing. There are some unused definitions that will
either be removed or become used in the follow up revisions.
We also obtained a PEN SMI number of 28218 from IANA for use
in MARF.

\subsection{SNMP Proxy Agents}

Based on the MIBs above, we produced proxy agents, using AdventNet's tools \cite{advent-net-java-agent-sdk}.
There are two kinds of proxies: one is proxy talking to the MARF's API and the agent, we call it {\em API proxy}
or {\em instrumentation proxy}.
The other one is proxy talking to manager and agents by using SNMP.
Because the time constraints, they are not fully instrumented.

Agents produced with the help of MIB Compiler of AdventNet also produces
two types of SNMP agents master agent (proxy to a group of SNMP sub-agents
that manager does not see) vs. sub-agents (SNMP) as well as
instrumentation (delegates, application-specific business logic). E.g. in
the case of Feature Extraction and LPC, the former would be the master
agent proxy, and LPC would be a sub-agent (e.g. a MIB subtree), which is
a more specific type of feature extraction.

\subsection{SNMP MARF Application Managers}

Within the timeframe of the course we did not manage to produce our own
SNMP manager application part and integrate it into the proposed GUI, so
we are leaving this task to the future work. The manager application we
used to test our work was the MIB Browser provided by the AdventNet's tools.

\subsection{Difficulties}

During our design, MIB validation, and compilation, we faced certain difficulties.
One of them the difference in ASN.1/SMI syntax checks between tools such as SimpleWeb vs AdventNet,
where we could not debug for long time one problem we faced: double AUGMENTS when we do the MIB of
general an concrete MARF services (e.g. Feature Extraction to LPC
or Classification to Neural Network, etc.).
MIB loading and configuration management in AdventNet and the corresponding
mapping of operations (instrumentation delegates in a pipeline) were a part
of the learning curve for AdventNet's API.

Here is the example that was holding us back where SimpleWeb's
validator \cite{simpleweb-mibvalidation} complained, but the
AdventNet's MIB Browser and Compiler didn't have problems with.
Assume:

\begin{verbatim}
tableFoo
tableEntryFoo

tableBar
tableEntryBar
   AUGMENTS {tableEntryFoo}
\end{verbatim}

\noindent
SimpleWeb's validator complained here:

\begin{verbatim}
tableBaz
tableEntryBaz
   AUGMENTS {tableEntryBar}
\end{verbatim}

\noindent
To give a more concrete example from one of our MIBs (a bit stripped).
In this case \api{lpcServiceEntry} would be managed by a sub-agent of
feature extraction. Please see more up-to-date MIB in \file{MARF-feature-extraction.mib}.

\small
\begin{verbatim}
  featureextractionServiceTable OBJECT-TYPE
      SYNTAX      SEQUENCE OF FeatureextractionServiceEntry
      MAX-ACCESS  not-accessible
      STATUS      current
      DESCRIPTION
          "The table of the Featureextraction services known by the SNMP agent."
      AUGMENTS { serviceTable }
      ::= { featureextractionService 1 }

  featureextractionServiceEntry OBJECT-TYPE
      SYNTAX      FeatureextractionServiceEntry
      MAX-ACCESS  not-accessible
      STATUS      current
      DESCRIPTION
          "Details about a particular Featureextraction service."
      AUGMENTS { serviceEntry }
      ::= { featureextractionServiceTable 1 }

  FeatureextractionServiceEntry ::= SEQUENCE {
      oFeatureSet     FeatureSet,
      adFeatures      VectorOfDoubles
  }

  lpcServiceTable OBJECT-TYPE
      SYNTAX SEQUENCE OF LPCServiceEntry
      MAX-ACCESS not-accessible
      STATUS current
      DESCRIPTION    "            "
      AUGMENTS { featureextractionServiceTable }
      ::={ featureextractionService 2 }

  lpcServiceEntry OBJECT-TYPE
      SYNTAX LPCServiceEntry
      MAX-ACCESS not-accessible
      STATUS current
      DESCRIPTION   "        "
      AUGMENTS { featureextractionServiceEntry }
      ::={ lpcServiceTable 1 }

  LPCServiceEntry ::=SEQUENCE {
      iPoles  INTEGER,
      iWindowLen INTEGER
  }
\end{verbatim}
\normalsize

\subsection{Contributions}

Generally, the entire team has focused on the development of
the MIBs, learning and trying out AdventNet's API, project presentation, and report equally.
Since MARF consists of multiple components, the workload was subdivided
roughly along those components, some common structures were worked
on together by the team. Some specific breakdown is as follows:

\begin{itemize}
\item
Serguei: overall project design and management, PEN application, MIB compilation, Classification and MARF server MIBs.
\item
Jian: learning MARF's API, Feature Extraction, LPC MIBs, AdventNet's MIB Compiler and API investigation, MIB diagrams.
\item
Lee Wei: learning MARF's API, Sample Loading, Preprocessing, Application MIBs, AdventNet's MIB Compiler and API investigation.
\end{itemize}

\subsection{Open Source}

This project, as the MARF itself, was developed as open-source project
as \url{SourceForge.net}. To checkout lastest version the source code, this report
sources, MIBs, and the presentation from our CVS repository, one can do by
executing the following commands:

\small
\begin{verbatim}
cvs -d:pserver:anonymous@marf.cvs.sourceforge.net:/cvsroot/marf login
cvs -z3 -d:pserver:anonymous@marf.cvs.sourceforge.net:/cvsroot/marf co -rINSE7120 -P marf
\end{verbatim}
\normalsize

\noindent
or alternatively browse it on-line:

\small
\begin{verbatim}
http://marf.cvs.sourceforge.net/marf/marf/?pathrev=INSE7120
\end{verbatim}
\normalsize

The files of the interest related to this project can be found
as follows: this report sources are in \file{marf/doc/src/tex/inse7120},
the presentation is in \file{marf/doc/presentations/inse7120}, the
corresponding graphics is in \file{marf/doc/src/graphics/distributed/mib},
the source code of the entire MARF is in \file{marf/src}, and the generated
agent code is in \file{marf/src/marf/net/snmp}.

\section{Future Work}
\label{sect:future-work}

This section summarizes future work highlights.

\subsection{Scenarios}

Here we present a few scenarios where DMARF along with the SNMP
management could be usefully employed. Management of these infrastructures
is better be conducted over SNMP, which is well adapted for the use of
networks under stress, along with RMON, and configuration management
by some kind of central authority.

\begin{itemize}
\item
Police agents in various law enforcement agencies spread out across a country,
yet, being able to identify speakers across all jurisdictions if say recorded phone
conversations of a suspect are available.
\item
Assume another scenario for conference microphones for different speakers
installed in a large room or separate rooms or even continents using
teleconferencing. We try to validate/identify who the speakers are at a
given point in time.
\item
Alternatively, say these are recorded voices of conference calls or phone
conversations, etc. We could have it reliable, distributed, with recovery
of agents/clients and the server database.
\item
Perhaps, other multimedia, VoIP, Skype, translation and interpretation natural
language services can be used with DMARF over the Internet and other
media.
\end{itemize}

\subsection{Summary}

There is a lot of distributed multimedia traffic and computation involved
between sample loading, preprocessing, feature extraction, and
classification servers. Efficiency in network management involves
avoidance of computations that already took place on another server and
just offloading the computed data over.

Since DMARF's originally implemented the Java RMI, SOAP (WebServices over XML-RPC),
and CORBA, some of that work can be transplanted towards the management needs.
In particular, DMARF's CORBA IDL definitions can as well be used for SNMP agent generation.
Thus, in this project another area to focus will be on the role of CORBA in network management
and the extension of Distributed MARF's \cite{dmarf06} CORBA services implementation
and its SpeakerIdentApp to provide efficient network management,
monitoring, multimedia (audio) transfer, fault-tolerance and recovery,
availability through replication, security, and configuration management.

More specifically, in the context of
MARF we'll look into few more of the aspects, based on the need, time, and interest.
Some research and implementation details to amend Distributed MARF we will consider
of the following:

\begin{itemize}
\item
Finish proxy agents and instrumentation.

\item
Implement our own managers
and the functions to compile new MIBs into the manager.

\item
Complete prototyped GUI
for ease-of-use of our management applications (as-is MARF is mostly console-based).

\item
Complete full statistics MIB and implement RMON along with
some performance management functions such as collecting
statistics and plotting the results.

\item
Propose a possible RFC.

\item
Make a public release and a publication.

\item
Implement some fault management functions such as alarms reporting.

\item
Look into XML in Network Management (possibly for XML-RPC).
\item
Look more in detail at Java and network management, JMX (right now through AdventNet).
\item
Distributed Management of different DMARF nodes from various locations.
\item
Management of Grid-based Computing in DMARF.
\item
Analysis of CORBA and where it fits in Network Management in DMARF.
\item
Multimedia Management using SNMP.
\end{itemize}

\section{Acknowledgments}

\begin{itemize}
\item Dr. Chadi Assi
\item SimpleWeb
\item Open-Source Community
\item AdventNet
\end{itemize}


	\addcontentsline{toc}{chapter}{Bibliography}

\bibliography{report}
\bibliographystyle{alpha}



	\printindex
\end{document}